\documentclass[reprint,aps]{revtex4-2}
\usepackage{amsmath,graphicx,float,amsfonts,cleveref,array,bm,soul}
 \usepackage[table,xcdraw]{xcolor}
\usepackage{algorithm}
\usepackage{algpseudocode}
\usepackage {pseudo}
\usepackage{appendix}
\usepackage{colortbl}
\usepackage{siunitx}
\usepackage{caption}
\usepackage{subcaption}
\usepackage{svg}

\begin{document}

\title{Analog, In-memory Compute Architectures for Artificial Intelligence}
\author{Patrick Bowen}
\email{ptbowen@neurophos.com}
\affiliation{Neurophos, 212 W Main St. \#301, Durham, North Carolina 27701, USA}
\affiliation{Center for Metamaterials and Integrated Plasmonics and Department of Electrical and Computer Engineering, Duke University, P.O. Box 90291, Durham, North Carolina 27708, USA}
\date{\today}

\author{Guy Regev}
\email{guy@alephzero.ai}
\affiliation{AlephZero, 5141 Beeman Ave. Valley Village, CA 91607, USA}
\affiliation{Department of Electrical and Computer Engineering, Ben-Gurion University of the Negev, David Ben-Gurion Blvd. 1, Be'er Sheva, Israel}
\date{\today}

\author{Nir Regev}
\email{nir@alephzero.ai}
\affiliation{AlephZero, 5141 Beeman Ave. Valley Village, CA 91607, USA}
\affiliation{Department of Electrical and Computer Engineering, Ben-Gurion University of the Negev, David Ben-Gurion Blvd. 1, Be'er Sheva, Israel}
\date{\today}
\author{Bruno Umbria Pedroni}
\email{bruno@alephzero.ai}
\affiliation{AlephZero, 5141 Beeman Ave., Valley Village, CA 91607, USA}
\affiliation{Department of Bioengineering, UC San Diego, 9500 Gilman Dr., La Jolla, CA 92093, USA}
\date{\today}

\author{Edward Hanson}
\email{edward.t.hanson@duke.edu}
\affiliation{Duke University, Electrical and Computer Engineering Department, Science Dr, Durham, NC 27710}

\author{Yiran Chen}
\email{yiran.chen@duke.edu}
\affiliation{Duke University, Electrical and Computer Engineering Department, Science Dr, Durham, NC 27710}

\begin{abstract}
    This paper presents an analysis of the fundamental limits on energy efficiency in both digital and analog in-memory computing architectures, and compares their performance to single instruction, single data (scalar) machines specifically in the context of machine inference.  The focus of the analysis is on how efficiency scales with the size, arithmetic intensity, and bit precision of the computation to be performed.  It is shown that analog, in-memory computing architectures can approach arbitrarily high energy efficiency as both the problem size and processor size scales.
\end{abstract}

\maketitle
\section{Introduction}
This work is focused on minimizing the energy required to evaluate neural networks, particularly in the linear layers which comprise the overwhelming majority of the computation.  The linear operators that describe convolutional neural network layers can be often be characterized by three qualities: they are sparse, high in dimensionality, and high in arithmetic intensity, where arithmetic intensity is defined as the ratio between the number of basic operations (i.e. multiplications and additions) and the number of bytes read and written. This paper shows that, in the context of operators that are both high in dimensionality and arithmetic intensity, an analog in-memory computing device can drastically reduce the energy required to evaluate the operator compared to a von Neumann  machine.  Moreover, the degree of increased efficiency of the analog processor is related to the scale of the processor.

In a classical von Neumann machine, the energy required to evaluate an operator can be broken into two components: memory access energy and computational energy. Within a typical CPU, and depending on the workload, these components can consume the same order of magnitude of the total energy. Memory access related energy can easily outgrow computational energy consumption, particularly when used to evaluate sequential large linear operators like those used in neural network inference.  The goal of this paper is to find high-level architectures that can reduce the energy consumption of neural network algorithms by orders of magnitude, which requires addressing both memory access energy and computational energy. Here we show that an in-memory compute accelerator architecture can reduce memory access energy when applied to an operator/algorithm with high arithmetic intensity, while an analog processor/accelerator can reduce computational energy when specialized for particular classes of linear operators. A processor architecture that takes advantage of both in-memory compute and is analog in nature can in principle reduce the overall computational energy consumption by orders of magnitude, with the amount of reduction depending on the scale and arithmetic intensity of the algorithm to be performed and the analog processor's specialization in performing a specific set of operators.

In-memory compute architectures were originally designed to speed up processing of algorithms that are parallelizable and applied to large datasets.  One of the earliest examples dates back to the 1960s with Westinghouse's Solomon project.  The goal of that project was to accelerate the speed of the computer up to 1 GFLOPs by using a single instruction applied to a large array of Arithmetic Logic Units (ALUs).  This is perhaps the first instance of the several closely related concepts: single instruction, multiple data (SIMD) machines, vector/array processors, systolic arrays and in-memory/near-memory compute devices.

Today, exploiting parallelism in high-arithmetic intensity algorithms using parallel hardware remains a well-known technique to accelerate a computation along the time dimension.  More recently, however, vector/array processors have been utilized to decrease compute energy as opposed to the original purpose of compute time, and it does this by reducing energy associated with memory accesses.  Google's TPU is a good example of a systolic array being used as a near-memory compute device with digital processing elements \cite{jouppi2017datacenter,samajdar2020systematic}.  In sec. \ref{sec:CIM}, we explain how in-memory compute devices can reduce memory access energy in the case of linear operators with high arithmetic intensity.

Separately, analog computing has recently been proposed as an approach to reduce the computational energy consumption, again for large, linear operations.  In sec. \ref{sec:AnalogComputing} we present a general model of analog computation that focuses on how energy consumption scales with problem size and bit precision, and show that computational energy can be reduced by orders of magnitude by using an analog processor that is specialized to implement specific classes of operators. Reconfigurable analog processors are by nature in-memory compute devices, and so these classes of processors are shown to reduce overall computational energy by orders of magnitude for particular operators.

\section{CPU Energy Consumption}
\label{sec:CPU}
We begin by finding the energy efficiency of a computer performing multiply-accumulate (MAC) operations, which are the core of linear operators used in deep learning.  The total energy required to perform a linear operation can be decomposed into memory access energy and computational energy:

\begin{equation}\label{eq:EnergyDecomposition}
    E_{tot}=N_{m}e_{m}+N_{op}e_{op},
\end{equation}
where $N_{m}$ is the number of memory accesses, $e_{m}$ is the average energy per access, $N_{op}$ is the number of operations required to evaluate the overall operator, and $e_{op}$ is the average energy per operation (e.g., add, multiply, etc).  We define the computational efficiency as the number of operations per unit energy performed by the computer:
\begin{equation}
    \eta\equiv N_{op}/E_{tot}=\frac{1}{(N_m/N_{op})e_m+e_{op}}.
\end{equation}

In a simple CPU with a single instruction, single data (SISD) architecture in Flynn's taxonomy and a flat memory hierarchy, for each operation that is performed, a value is read from memory for the current partial sum, the operator weight, and the input activation.  The three values are operated upon, and the result is written back to memory.  Therefore, regardless of the actual size of the weights or activations, the number of memory accesses per operation will always be four (i.e. three reads and one write), and the number of computational operations (multiply and add) will be 2. This results in $N_m=2N_{op}$ and a computational efficiency of
\begin{equation}\label{eq:CPU_Efficiency}
    \eta=\frac{1}{2e_{m}+e_{op}}.
\end{equation}

In modern CMOS devices, both $e_m$ and $e_{op}$ are on the order of magnitude of 1 pJ \cite{horowitz20141}, as will later be shown in \cref{tab:EnergyPerOperation}.  This places an approximate limit on the computational efficiency of most traditional architectures on the order of 0.1-1 TOPS/W, which is consistent with state of the art performance \cite{reuther2019survey}.

\section{Minimizing Memory Access Energy with in-memory compute}
\label{sec:CIM}
One of the major downsides of SISD machines is that they can end up accessing the same memory element multiple times in the course of evaluating a large operator, which wastes memory access energy.  This is ultimately reflected in the ratio $N_{op}/N_m=1/2$ that is fixed by the nature of a SISD machine. Alternatively, one can imagine finding another hypothetical architecture that is arranged in some energetically optimal way to where all of the inputs are only read once from memory, and all outputs are only written once to memory in the course of the computation.  If that were done, this would represent the minimum total access energy required to evaluate the linear operator.  In other words, $N_m$ would reach its minimum value, and the ratio $N_{op}/N_m$ would be maximized.

\begin{figure}
    \centering
    \includegraphics[width=0.4\textwidth,trim=0 6cm 0 6cm]{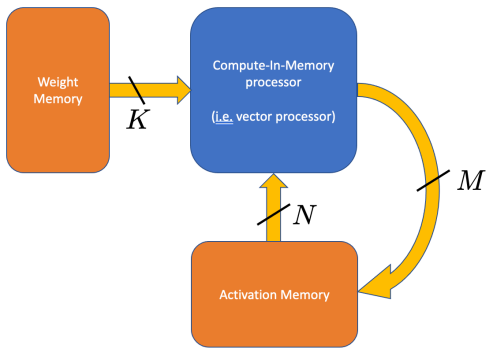}
    \caption{Illustration of a digital compute-in-memory processor.}
    \label{fig:IMC_Device}
\end{figure}

While a particular processor might only be able to implement a certain $N_{op}/N_m$ ratio, this ratio is also limited by the algorithm being performed, and is commonly referred to as the \emph{arithmetic intensity} of the algorithm:
\begin{equation}
    a\equiv N_{op}/N_m.
\end{equation}
An \emph{in-memory compute} device \cite{sebastian2020memory} as illustrated in \cref{fig:IMC_Device} can leverage the arithmetic intensity of an algorithm by reading a large set of both operator data and input vector data from memory at once and operating on all of the data together before writing the output back to memory.  If the in-memory compute device is sufficiently large and complex, all of the necessary operations involving this data can be performed without any of the inputs being read a second time from memory in the future.

Returning to \cref{eq:EnergyDecomposition}, we set a lower bound on the amount of memory access energy that must be expended for the von Neumann machine to evaluate the operator in terms of the arithmetic intensity.  This in turn leads to a limit on the computational efficiency:
\begin{equation}
\label{eq:DigitalCIMEfficiency}
    \eta=\frac{1}{e_{m}/a+e_{op}}
\end{equation}
The contribution to computational efficiency from memory access energy can therefore be brought arbitrarily low when implementing an operator with arbitrarily high arithmetic intensity.  The reduction in the contribution from memory access energy with increasing arithmetic intensity in \cref{eq:DigitalCIMEfficiency} is reflective of the energy savings in systolic arrays and TPUs \cite{jouppi2017datacenter,samajdar2020systematic}. 

We note that the kind of analysis presented in \cref{eq:DigitalCIMEfficiency} is analogous to roofline models of processors \cite{williams2009roofline}; however, the emphasis here is on energy consumption, while the latter is focused on identifying bottlenecks in processor speed.

In order to sample what degree of advantage in-memory compute devices can bring, we examine a few examples of linear operators and present their arithmetic intensities.  For a general matrix multiplication of a matrix of size $L\times N$ times a matrix of dimension $N\times M$ the total number of memory accesses is $N_m=LN+NM+LM$, and the number of operations is $N_{op}=2NML$, where additions and multiplications are treated as separate operations. The arithmetic intensity in this case is:
\begin{equation}\label{eq:computeEfficieny1}
    a=\frac{2NML}{LN+NM+LM},
\end{equation}
which approaches $\infty$ as $N,M,L\rightarrow\infty$ collectively.  

\begin{center}
\begin{figure*}
    \hspace*{-3.5cm}
    \centering
    \includegraphics[width=0.7\textwidth,trim=0cm 0cm 0cm 0cm]{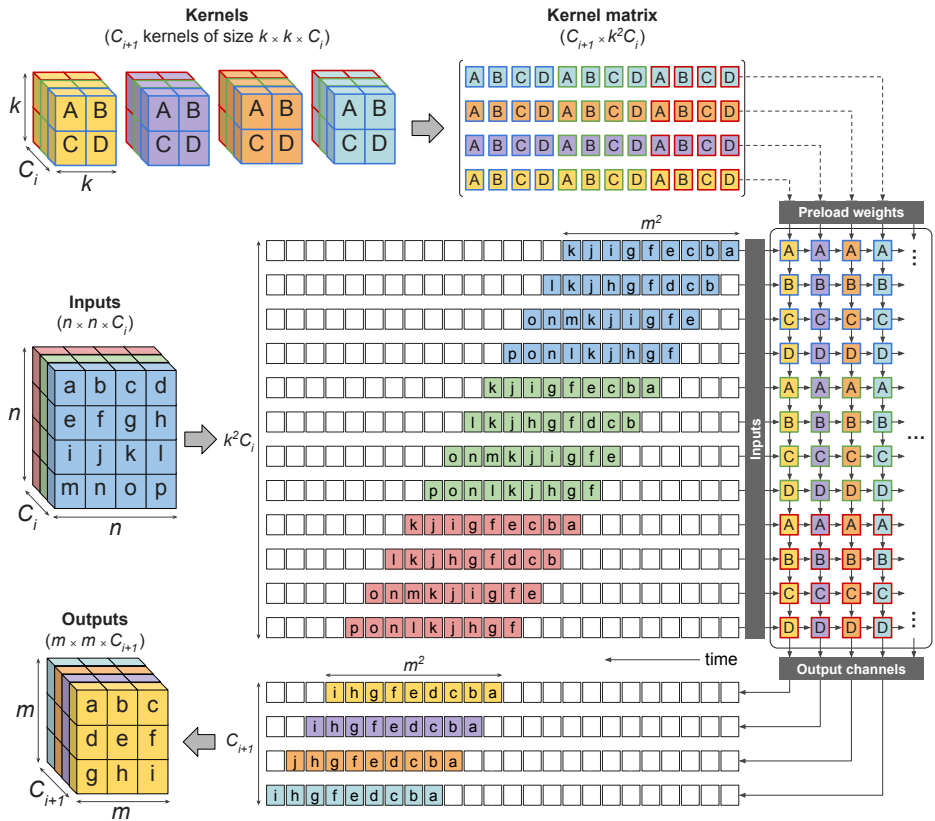}
    \caption{Algorithmic implementation of a convolution using matrix multiplication in a weight-stationary systolic array.  The input data is converted into a toepliz matrix and fed into the systolic array, with each row delayed one time step behind the one above it.}
    \label{fig:im2col}
\end{figure*}
\end{center}

For a convolution, the arithmetic intensity can similarly become arbitrarily large, since a convolution can be implemented as a matrix-matrix multiplication.  This is typically done by rearranging the input data into a toeplitz matrix using what is known as an im2col() operation.  The general algorithm of implementing convolution using matrix multiplication in a systolic array is shown in \cref{fig:im2col}, where $n \times n$ is the size of one input channel, $C_i$ is the number of input channels, $k \times k$ is the size of one of the kernel channels, and $C_{i+1}$ is the number of output channels (and, consequently, also the number of individual 3-D kernels). The toeplitz formed by replicating and rearranging the activation data results in an $(n-k+1)^2 \times k^2C_i$ matrix.  A convolution is performed by multiplying this with a $k^2C_i\times C_{i+1}$ matrix containing the weights.  Therefore, when implementing a convolution using matrix multiplication we generally have matrix dimensions,
\begin{subequations}
\begin{align}\label{eq:AnalogProcessorDimensions}
    L&= (n-k+1)^2\approx n^2\\
    N&= k^2C_i\\
    M&= C_{i+1}.\label{eq:AnalogOutputDimension}
\end{align}
\end{subequations}
which results in an arithmetic intensity,
\begin{equation}\label{eq:ConvArithmeticIntensity0}
    a=\frac{2n^2k^2C_iC_{i+1}}{n^2k^2C_i+k^2C_iC_{i+1}+n^2C_{i+1}}.
\end{equation}

However, since the activation data was replicated approximately $k^2$ times in order to form the input matrix, the arithmetic intensity is significantly reduced relative to a processor that natively implements convolution instead of general matrix multiplication.  To see this, consider again the convolutional layer of an $n\times n$ input image  with $C_i$ input channels, $C_{i+1}$ output channels, and a $k\times k$ kernel.  The input vector size is $N_i=n^2C_i$, and the number of kernel weights is $K=k^2C_iC_{i+1}$.  If only the necessary weight and activation data were required to be read, the arithmetic intensity of the $i^{th}$ layer would become
\begin{equation}\label{eq:ConvArithmeticIntensity}
    a\approx\frac{2n^2k^2C_iC_{i+1}}{n^2(C_i+C_{i+1})+k^2C_iC_{i+1}}.
\end{equation}
In the limit where $n^2>>k^2C_i$, this is roughly $k^2$ higher arithmetic intensity than when convolution is implemented using matrix multiplication.

Whether convolution is implemented natively or using matrix-matrix multiplication, \cref{eq:ConvArithmeticIntensity} shows that, as $n,k,C_{i}\rightarrow\infty$, arithmetic intensity becomes arbitrarily large, making the contribution from memory access energy in \cref{eq:DigitalCIMEfficiency} arbitrarily small.  Indeed, in most modern convolutional neural networks, these parameters are large and yield high arithmetic intensity, as shown in \cref{tab:CNN_Params}.  Depending on the size of the memory banks (which determine memory access energy), and based on the reference numbers given in \cref{tab:EnergyPerOperation} for SRAM access energy and digital MAC operation, an in-memory compute processor implementing an algorithm with high arithmetic intensity can be made to expend negligible memory access energy relative to the computational energy.

\begin{table*}[]
\begin{tabular}{|p{0.155\textwidth}|>{\centering}p{0.1\textwidth}|>{\centering}p{0.09\textwidth}|>{\centering}p{0.1\textwidth}|>{\centering}p{0.09\textwidth}|>{\centering}p{0.07\textwidth}|>{\centering\arraybackslash}p{0.09\textwidth}|>{\centering\arraybackslash}p{0.12\textwidth}|>{\centering\arraybackslash}p{0.09\textwidth}|}
\hline
\textbf{Network} & \textbf{\# of layers} & \textbf{median $n$} & \textbf{median $C_i$} & \textbf{max $N$} & \textbf{avg. $k$} & \textbf{total $K$} & \textbf{median $C_{i+1}$} & \textbf{median $a$} \\ \hline
DenseNet201       &        200 &      62 &     128 & 1.6e+07 &       2.0 & 1.8e+07 &     128 &  292 \\ \hline
GoogLeNet         &         59 &      61 &     480 & 3.9e+06 &       2.1 & 6.1e+06 &     128 &  200 \\ \hline
InceptionResNetV2 &        244 &      60 &     320 & 8.0e+06 &       1.9 & 8.0e+07 &     192 &  291 \\ \hline
InceptionV3       &         94 &      60 &     192 & 8.0e+06 &       2.4 & 3.7e+07 &     192 &  295 \\ \hline
ResNet152         &        155 &      63 &     256 & 1.6e+07 &       1.7 & 5.8e+07 &     256 &  390 \\ \hline
VGG16             &         13 &     249 &     256 & 6.4e+07 &       3.0 & 1.5e+07 &     256 & 2262 \\ \hline
VGG19             &         16 &     186 &     256 & 6.4e+07 &       3.0 & 2.0e+07 &     384 & 2527 \\ \hline
YOLOv3            &         75 &      62 &     256 & 3.2e+07 &       2.0 & 6.2e+07 &     256 &  504 \\ \hline
\end{tabular}
\caption{Summary of convolutional layer parameters of various well-known neural networks considering a 1-Mpixel (per channel) input image.}\label{tab:CNN_Params}
\end{table*}

\section{Reducing Computational Energy with Analog Computing}
\label{sec:AnalogComputing}

\begin{figure*}
    \centering
    \includegraphics[width=\textwidth]{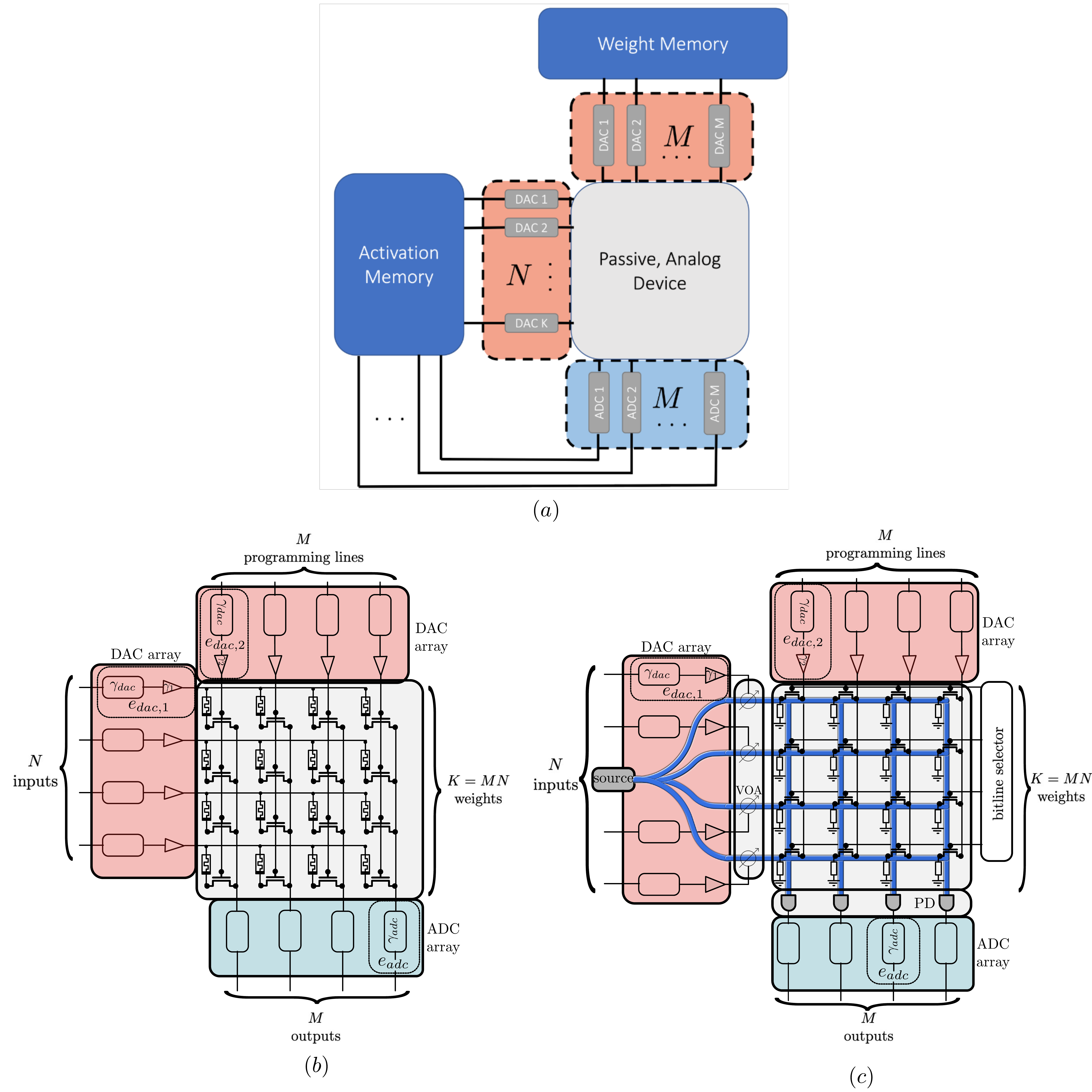}
    \caption{(a) System-level view analog, in-memory compute processors. The analog device is configured using DACs to either hold activations or weights, while the other is provided as input. (b) Detailed view of a ReRAM crossbar analog electronic in-memory compute processor.  Each transistor is connected to a reconfigurable resistor, the conductance of which determines the effective weight of each element in the matrix.  (c) Detailed view of a silicon photonic in-memory compute processor.  Each transistor is connected to an electro-optic element that changes the scattering parameters through each intersection.}
    \label{fig:CIM_Analog}
\end{figure*}

Unfortunately, by Ahmdal's law, even if the memory access energy is made arbitrarily small, computational energy consumed by the logical units will limit the overall performance gains to be made.  In order to improve the overall efficiency by orders of magnitude, both contributions need to be addressed.

Recently, various types of analog computing, from electrical to optical, have been proposed as techniques to reduce computational energy consumption.  Electronic analog computing typically centers around crossbar arrays of resistive memory (or ReRAM) \cite{demler2018mythic,xia2019memristive,shafiee2016isaac}. Optical analog processors are commonly based on silicon photonics \cite{feldmann2021parallel,ong2020photonic,fang2019design,shen2017deep}.  Optical 4F systems have been explored since the 1980s as a higher dimensional form of compute \cite{ambs2010optical,chang2018hybrid}, and simple scattering off of optical surfaces is also being explored \cite{lin2018all,qian2020performing,zhou2021large}.

The argument for analog computing is fundamentally a scaling one: analog computing has particular advantages when applied to large, linear operators with low bit precision \cite{hamerly2019large}.  To see this, consider a general analog processor (shown in \cref{fig:CIM_Analog}(a)) that takes $N$ numbers of $B$-bit precision input data, produces $M$ numbers of $B$-bit precision output data, and is configured by $K$ weights with $B$-bit precision which represent the matrix. The analog processor is first configured by converting the $K$ weights using digital-to-analog converters (DACs) and applying these values to the modulators in the analog processor.  Then the $N$ inputs are read from memory, and DACs are used to apply $N$ analog inputs to the processor.  By the physics of the processor, this naturally results in $M$ analog outputs, which are converted back to the digital domain using analog-to-digital (DAC) converters.  If the analog processor is somehow already configured, or never needs to be reconfigured, then the total energy consumed will be only that of the DACs for the inputs and ADCs for the outputs:

\begin{equation}\label{eq:AnalogEnergy}
    E_{op} \equiv N_{op}e_{op}=N(e_{dac,1}+e_{adc}),
\end{equation}
where we have assumed $N=M$ for simplicity.  While the right-hand-side of \cref{eq:AnalogEnergy} represents the computational energy consumed by the analog processor, the left-hand-side represents the equivalent number of digital operations performed ($N_{op}$) times the energy that each of those operations would have to take ($e_{op}$) in order for a digital computer to achieve the same efficiency as the analog computer.  Since $N_{op}=2N^2$ for matrix multiplication, if this operation were performed digitally, the expended computational energy would be proportional to the number of operations: $E_{op}=2e_{op}N^2$.  The conclusion is that \emph{analog computing reduces matrix multiplication from $\mathcal{O}(N^2)$ in energy to $\mathcal{O}(N)$ in energy}.  This furthermore implies that the effective energy per operation of analog computing scales inversely to the size of the problem, i.e.
\begin{equation}\label{eq:AnalogScaling}
    e_{op}\propto 1/N.
\end{equation}
We note that in practice the scaling $N$ is defined either by the size of the processor or the size of the problem, whichever is smaller.

\subsection{Vector-Matrix Multiplication}
For most problems involving neural networks, the analog processors that can be created are not large enough to store the entire neural network.  In this case, the reconfiguring of the weights in the analog processor itself can destroy the $\mathcal{O}(N)$ scaling advantage.  To see this, consider the multiplication of a vector of length $N$ with a matrix of dimensions $N\times M$.  In this case,  we have,
\begin{align}
    N_{op}e_{op}&=2Ne_{dac,1}+2MNe_{dac,2}+2Me_{adc}.
    \label{eq:ComputeEnergy}
\end{align}
We have also separated the DAC energies $e_{dac,1}$ and $e_{dac,2}$ since different physical mechanisms and loads are sometimes used to configure an analog computer versus feed it with analog inputs.  Here, $e_{dac,1}$ is used to represent the energy required per input, while $e_{dac,2}$ is used to represent the energy required per reconfiguration.  

Typically, in analog computing technologies, the analog in-memory compute device can only store either positive definite numbers (like in the example of memristors) or fully complex numbers (like in the case of coupled Mach-Zender interferometers).  If only positive numbers can be created, then the entire calculation must be done twice and the difference of the results taken in order to take into account both positive and negative matrix values.  On the other hand, when complex values are allowed like in the case of silicon photonic MZI's, there are two voltages (and hence two DAC operations) required to configure each coupled MZI modulator.  Additionally, for coherent optical measurements, an interference technique must be used to recover the positive and negative field components from the photodetectors, which can only measure the norm square of the field.  Hence, regardless of the analog compute scheme, each term in \cref{eq:ComputeEnergy} must practically be multiplied by a factor of two in order to handle both positive and negative values.

Applying \cref{eq:ComputeEnergy} to vector-matrix multiplication, we obtain:
\begin{equation}
    e_{op}=e_{dac,1}/M+e_{dac,2}+e_{adc}/N,
    \label{eq:VM_ComputeEnergy}
\end{equation}
in which case the middle term is proportional neither to $1/N$ nor $1/M$.  

\subsection{Matrix-Matrix Multiplication}
The aforementioned situation is relieved in the case of  matrix-matrix  multiplication.  In this case the configuration of the analog computer itself is reused for every row of the input matrix, restoring the energy cost per operation to be inversely proportional to the problem scaling.  In the case of an $L\times N$ matrix times an $N\times M$ matrix, we have
\begin{equation}\label{eq:MM_ComputeEnergy}
    e_{op}=e_{dac,1}/M+e_{dac,2}/L+e_{adc}/N
\end{equation}
since $N_{op}=2NML$ in this case.  Since each of the three separate contributions to the energy consumption is decreased by a factor proportional to the three different dimensions associated with the matrices being multiplied, the effective energy per operation decreases as the problem scale increases.  In the case of a finite-sized analog processor, the last two contributions will ultimately be limited by the two dimensions (number of inputs and outputs) of the analog processor itself.

At this point, a distinction needs to be made between the size of the matrices involved in the neural net architecture and the physical dimensions of the analog processor.  We label the matrix dimensions with primes, i.e. $M'$, $N'$, and $L'$, and label the physical dimensions of the processor with hats: $\hat{M}$, $\hat{N}$.  The actual factors by which energy is saved (i.e. $M$ and $N$ in \cref{eq:MM_ComputeEnergy}) are given by the smaller of these two numbers:
\begin{subequations}
\begin{align}
    M&=\mathrm{min}\{\hat{M},M'\} \\
    N&=\mathrm{min}\{\hat{N},N'\}.
\end{align}
\end{subequations}

\subsection{Convolution}
As in the case of digital processors, analog processors can also implement convolution using matrix-matrix multiplication.   The mapping of the kernel and activation data to to matrix dimensions remains the same, i.e. 
\begin{subequations}
\begin{align}
    L'&= (n-k+1)^2\approx n^2\\
    N'&= k^2C_i\\
    M'&= C_{i+1}\label{eq:AnalogOutputDimension}
\end{align}
\label{eq:AnalogProcessorDimensions}
\end{subequations}
when weight-stationary scheme is implemented.  These numbers are permuted for activation-stationary.  
As with digital processors, one of the unfortunate aspects of representing convolution as pure matrix multiplication is that the input activations get duplicated $k^2$ times, which means $k^2$ more DAC operations (and possibly memory accesses as well) than in a processor that natively implements convolution rather than general matrix multiplication.  The consequence of this is that $M$ is by far the smallest of the numbers in \cref{eq:AnalogOutputDimension}, and therefore analog processors that implement convolution as matrix multiplication get the least amortization over their input DACs in \cref{eq:MM_ComputeEnergy}. The median values of $L'$, $N'$, and $M'$ for various neural networks is presented in \cref{tab:LNM_Analog}.

\begin{table}[]
\begin{tabular}{|l|c|c|c|c|} \hline
\textbf{Network} & \textbf{\# of layers} & \textbf{$L'$} & \textbf{$N'$} & \textbf{$M'$} \\ \hline
DenseNet201        & 200 & 3844  & 1152 & 128  \\ \hline
GoogLeNet          & 59  & 3721  & 528  & 128  \\ \hline
InceptionResNetV2  & 244 & 3600  & 432  & 192  \\ \hline
InceptionV3        & 94  & 3600  & 768  & 192  \\ \hline
ResNet152          & 155 & 3969  & 1024 & 256  \\ \hline
VGG16              & 13  & 62001 & 2304 & 256  \\ \hline
VGG19              & 16  & 38688 & 2304 & 384  \\ \hline
YOLOv3             & 75  & 3844  & 1024 & 256  \\ \hline
\end{tabular}\\
\caption{Median values of $L'$, $N'$, and $M'$ as per \cref{eq:AnalogProcessorDimensions} for the convolutional layers of various well-known neural networks. The values were obtained considering a 1-Mpixel (per channel) input image.}\label{tab:LNM_Analog}
\end{table}

\section{Operator-Specialized Analog Processors}
Thus far, we have seen that 1) the contribution of memory access energy to compute efficiency can be brought arbitrarily low by implementing networks with large arithmetic intensity on specialized processors, and 2) analog processors can further reduce computational energy consumption when performing matrix multiplication.  The reduction in computational energy is proportional to the size of the matrix the analog processor can handle.

One of the inherent disadvantages of planar, matrix multiplication based processors in performing convolutions is that the matrix that is formed for the input is of dimensions $(n-k+1)^2\times k^2 C_i$, which is a factor of $k^2$ larger than the actual activation data.  When the convolution is performed digitally this is of little consequence because the number of MACs required is the same for this matrix multiplication as it is for convolution: $(n-k+1)^2k^2C_i$.  However, when the matrix multiplication is performed with an analog processor using a matrix with $k^2$ more rows than necessary means that it requires $k^2$ more DAC operations than should be theoretically necessary.  Even worse than this, unless some additional logic is used to set up the matrix between the SRAM and processor (which also consumes energy), it will require $k^2$ additional memory reads than is in principle necessary, thus significantly increasing the memory access energy.  Furthermore, since the number of channels in each layer are often correlated (the output channels of one layer become the input channels of the next), the weight data loaded into the analog processor which has dimensions of $k^2C_i\times C_{i+1}$ is highly rectangular, which will increase $M$ relative to $N$, which in turn increases the contribution to the energy consumption per operation to the input data DACs.

In contrast to analog processors designed for general matrix multiplication, there are classes of analog processors which are specialized to implementing convolutions.  One technique to implementing an analog processor is by restricting it to only operate in one particular eigenspace of operators.  While any linear operator may be expressed as a matrix, the matrix $A$ may be factored into the product of three matrices using eigen-decomposition:
\begin{equation}
    X=U\Lambda U^T,
\end{equation}
where $U$ is a unitary (i.e. lossless) matrix of the eigenvectors of $X$, and $\Lambda$ is a purely diagonal matrix of the eigenvalues of $X$.  The eigenvectors of a convolution are waves, and so when $X$ is a matrix representing a convolution, the eigenvector matrix $U$ represents a Fourier transform, while $U^T$ represents and inverse Fourier transform.

\begin{figure*}
    \centering
    \includegraphics[width=\textwidth]{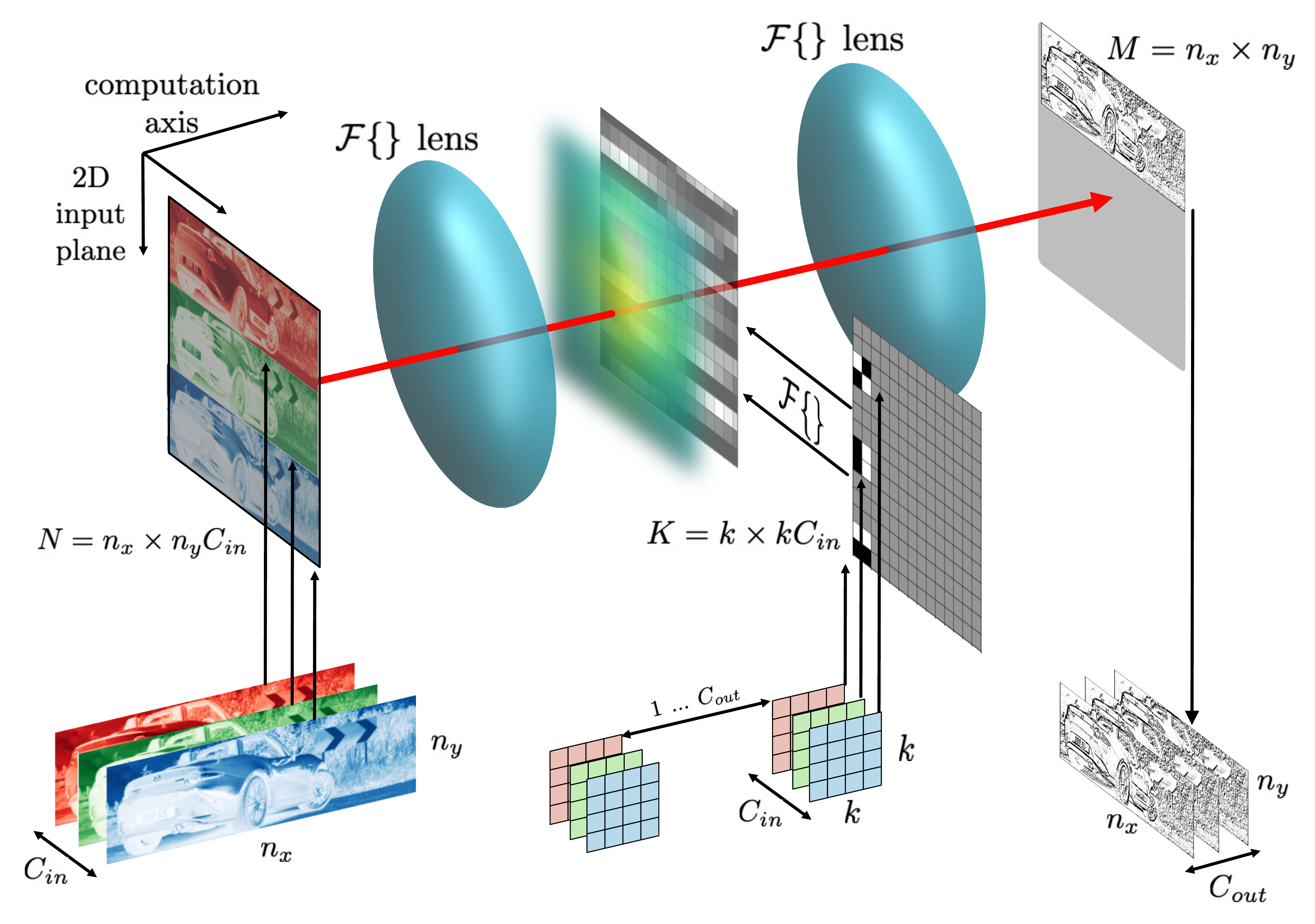}
    \caption{Illustration of a transmission-mode optical 4F system performing convolutions with parallelized input channels.  The input activation data can be tiled on the object plane, while the input filters can be tiled with appropriate padding before the Fourier transform is taken and the data is applied to the second SLM in the Fourier plane.  In this arrangement one complete output channel is produced per measurement.}
    \label{fig:4F_figure}
\end{figure*}

One technique of creating an \emph{operator-specialized processor} is to statically implement the matrices $U$ and $U^T$, and only dynamically reconfigure the eigenvalues $\Lambda$.  In this case, in order to change linear operators from one to another only the diagonal entries of $\Lambda$ need to be changed.  In other words, if the matrix $X$ is of size $m\times m$, changing the matrix to another convolution matrix only requires the modulation of $m$ weights in the analog processor instead of $m^2$ weights.  In the particular case where $X$ represents a convolution, these eigenvalues are the Fourier transform of the kernel data.  By tuning this set of $m$ elements, the matrix $X$ that is implemented by the analog processor can span the range of linear operators with the eigenvectors given by $U$.

Eigen-decomposition is possible for planar analog processors, and has in fact been demonstrated in silicon photonic processors \cite{ong2020photonic,shen2017deep}. However, there is an alternate approach to silicon photonics to implementing a convolution-specialized processor called an \emph{optical 4F system}, which has a particular set of advantages relative to planar convolution processors.  

In planar analog processors, data is inserted into the processor in a one dimensional array, and the data is processed as it propagates along the second dimension.  Unlike planar processors, an optical 4F system is a volumetric processor, so data is represented in a two dimensional array, while the computation happens as light propagates in the third dimension.  While this does bring dramatically higher information density and computational density, the most significant difference is that it allows the processor to scale to numbers of inputs that are entirely impractical for planar processors.  Since the efficiency of analog compute was shown in \cref{eq:AnalogScaling} to scale proportionally to the dimensions of the analog processor (in the limit of infinite arithmetic intensity), optical 4F systems can in theory reach computational efficiencies orders of magnitude higher than planar processors.

An example of an optical 4F system processor is shown in \cref{fig:4F_figure}. It is composed of two spatial light modulators (SLMs), which might be based on either liquid crystal cells or dynamic metasurfaces.  These are placed before and after a lens, one focal length away from either side.  Lenses naturally perform Fourier transforms between these two place, so that the light transmitted through the first SLM is Fourier-transformed upon passing through the lens.  Therefore the first SLM provides the input data, and the first lens represents multiplication by the unitary Fourier matrix $U$.  The second SLM is loaded with the Fourier transform of the kernel data and the transmitted light through it; therefore, the product of the Fourier transform of the input data with the Fourier transform of the kernel data.  The second SLM therefore represents the multiplication by the diagonal eigenvalue matrix $\Lambda$.

A second lens is then placed after the second SLM, one focal length away, which represents multiplication by the second eigenvalue matrix $U$.  Finally, a detector is placed a second focal length from the second lens, and the light impinging on the detector is therefore the convolution of the input data with the kernel data.  The detector itself is sensitive only to the intensity (i.e. the norm square) of the incident field.  However, the complex value of the field can nonetheless be recovered using interferometric methods.  Alternatively, as others have pointed out, the nonlinear measurement performed by the optical absorption of semiconductors can also be used naturally as the nonlinear activation of the neurons.

As shown in \cref{fig:4F_figure}, more than one input channel can be processed in parallel if the kernel data is appropriately padded before the Fourier transform is taken and the data is applied to the second SLM.  This allows greater SLM utilization when small kernels are being used.

\begin{figure*}
    \centering
    \includegraphics[width=\textwidth]{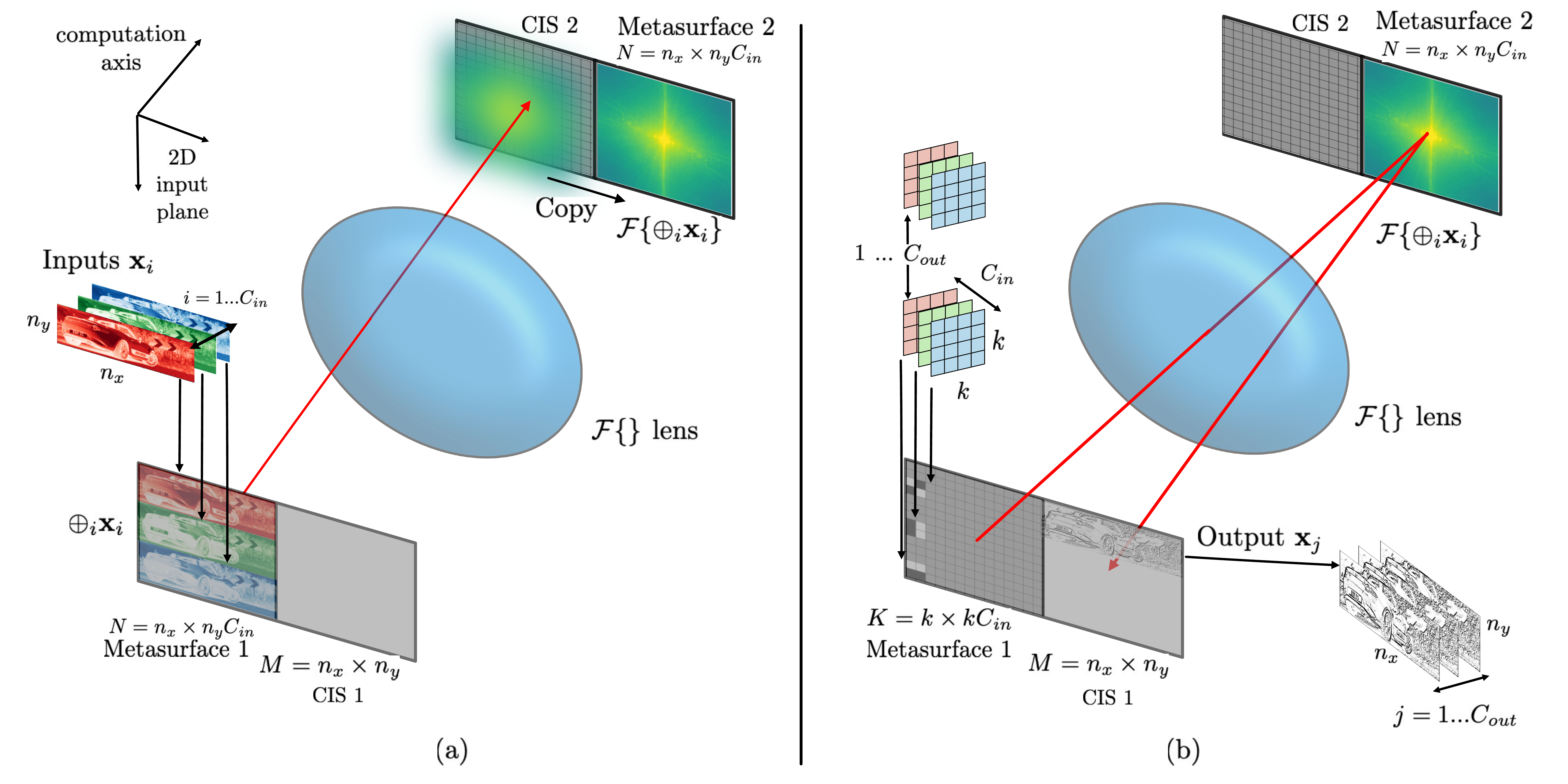}
    \caption{Illustration of a reflection-mode optical 4F system (folded into a 2F overall length) performing processing a full convolutional layer with all input and output channels in two phases: (a) phase one, where an optical Fourier transform of the input activation data is taken and loaded into the Fourier plane SLM, (b) phase two, where the input channels are tiled onto the object plane SLM and the convolution of all input channels are measured in parallel.  The process is repeated for each output channel.}
    \label{fig:In-Reflection2}
\end{figure*}

Unfortunately, from a compute systems perspective, traditional optical 4F systems have a fatal flaw: the output data from the convolution is measured four focal lengths away from the input data, which presumably must be physically implemented in its own chip.  Since this convolution operation only represents the connections between two layers of neurons, in order to implement a deep neural network with more than two layers of neurons the output data from the detector chip must be brought back somehow to the input spatial light modulator.  Communicating this massive amount of data off-chip would entail massive energy costs, overcoming all advantages brought by the large-scale analog compute.

However, an optical 4F system might be folded using reflection-mode SLMs as shown in \cref{fig:In-Reflection2} in order to consolidate the first SLM and the CMOS image sensor side-by-side into a single chip, and using only a single lens.  In this architecture all significant data transfer between the two chips happens optically instead of electronically.  On either side of the lens on two chips, split into two halves are: an SLM (or metasurface) and a CMOS image sensor.  Both chips are placed one focal length away from either side of the lens such that, whenever light passes between the two chips, a Fourier transform is taken by the lens.

This system computes convolutions in two phases: a loading phase and a compute phase.  The first, loading phase is shown in \cref{fig:In-Reflection2}(a), where the purpose is to take the Fourier transform of the activation data and load it into the second metasurface.  A set of input filter maps are written to the input SLM in the first chip, which is illuminated.  The Fourier transform of the reflected light is delivered to the CMOS image sensor (CIS) in the second chip, and this data is electronically transferred over to the second SLM within the same chip using DAC and ADC operations.  As with the in-transmission unfolded 4F system in \cref{fig:4F_figure}, in-reflection 4F systems like the one in \cref{fig:In-Reflection2} can be used to take the convolution of multiple input channels in parallel.  The final result of this phase is therefore that the SLM in the second chip is configured with the Fourier transform of the activation data.  

In the second, compute phase, the input kernel weight data is applied to the first SLM. This is then illuminated at a slightly oblique angle so that the reflected light impinges upon the SLM in the second chip.  When this light is reflected the lens takes another Fourier transform, and the light impinging on the CIS in the first chip is the convolution of the input filter map data with the kernel data.  

If the input data requires $n^2C_i$ total pixels, loading the optical Fourier transform of the activation data will cost
\begin{equation}
    E_{fft}=n^2C_i(2e_{adc}+4e_{dac})
\end{equation}
energy.  One DAC operation per pixel is required to write the input data to the first metasurface, while two ADC operations and two DAC operations are required in order to reconstruct the complex field data from the intensity data and then apply it to the second SLM.

Since input channels can be performed in parallel and then looped over output channels, the second phase involves two times $K=k^2C_iC_{i+1}$ DAC operations, and two $n^2C_{i+1}$ ADC operations in the CIS to recover the field.  
\begin{equation}
    E_{conv}=2k^2C_iC_{i+1}e_{dac}+2n^2C_{i+1}e_{adc}
\end{equation}

Therefore the total energy associated with the analog compute of this layer is $E_{fft}+E_{conv}$,
\begin{equation}
    E_{op}=2n^2(C_i+C_{i+1})e_{adc}+2C_i(2n^2+k^2C_{i+1})e_{dac}.
\end{equation}
The total number of operations performed is $N_{op}=2n^2k^2C_iC_{i+1}$.  Therefore the efficiency of the approach is,
\begin{equation}
    \eta=\frac{1}{e_m/a+e_{adc}/\left(\frac{k^2C_iC_{i+1}}{C_i+C_{i+1}}\right)+2e_{dac}/k^2C_{i+1}+e_{dac}/n^2}.
\end{equation}
in the limit that the metasurfaces are large enough to handle all of the activation or weight data.

In order to take into account the finite size of the metasurfaces, which may not be large enough to fit all of the activation data from all channels at once, we first find the number of input channels that can practically be handled at once.  For a metasurface of dimension $n_x\times n_y\equiv \hat{N}$, the number of input channels that can be included at once, $C'$, is,
\begin{equation}
    C'=\lfloor\hat{N}/n^2\rfloor.
\end{equation}

Using this in place of the actual number of software defined input channels we can derive the factors by which energy is saved in the optical 4F system in the case that $C'\geq1$,
\begin{subequations}
\begin{align}
    L&=n^2 \\
    N&=\frac{k^2C'C_{i+1}}{(C'+C_{i+1})}\\
    M&= k^2C_{i+1}/2.
\end{align}
\label{eq:4F_Dimensions}
\end{subequations}
In terms of these parameters, the efficiency of the optical 4F system is given in the usual way,
\begin{equation}\label{eq:4F_Efficiency}
    e_{op}=e_{dac}/M+e_{dac}/L+e_{adc}/N.
\end{equation}

\noindent For an optical 4F system, the median values of $L$, $N$, and $M$ as per \cref{eq:4F_Dimensions} for various neural networks is presented in \cref{tab:LNM_4F}.

\begin{table}[]
\begin{tabular}{|l|c|c|c|c|} \hline
\textbf{Network} & \textbf{\# of layers} & \textbf{$L$} & \textbf{$N$} & \textbf{$M$} \\ \hline
DenseNet201        & 200 & 3844  & 272  & 136   \\ \hline
GoogLeNet          & 59  & 3721  & 128  & 64    \\ \hline
InceptionResNetV2  & 244 & 3600  & 224  & 112   \\ \hline
InceptionV3        & 94  & 3600  & 240  & 120   \\ \hline
ResNet152          & 155 & 3969  & 1024 & 512   \\ \hline
VGG16              & 13  & 62001 & 2304 & 1152  \\ \hline
VGG19              & 16  & 38688 & 3456 & 1728  \\ \hline
YOLOv3             & 75  & 3844  & 512  & 256   \\ \hline
\end{tabular}\\
\caption{Median values of $L$, $N$, and $M$ for the convolutional layers of various well-known neural networks considering an optical 4F system as computational substrate. The values were obtained considering a 1-Mpixel (per channel) input image and an infinitely large metasurface (i.e. $C'\rightarrow\inf$).}\label{tab:LNM_4F}
\end{table}

\section{Analytic Results}
The formula given in \cref{eq:CPU_Efficiency,eq:DigitalCIMEfficiency,eq:MM_ComputeEnergy,eq:4F_Efficiency} can be used to estimate the efficiency when evaluating a given CNN layer on any one of those four compute platforms.  They depend on the energy values for memory access, DAC/ADC operations, and digital multiplication.  Estimates for many of these quantities are given in \cref{tab:EnergyPerOperation}, and formula for deriving the loads to estimate DAC energies for various analog compute platforms are also given in the appendix.  

\begin{table}[]
    \centering
    \begin{tabular}{|c|c|}
         \hline $e_m$ (96kB SRAM)\cite{horowitz20141} & 4.3pJ\\
         \hline $e_{mac}$ \cite{horowitz20141}& 0.23pJ\\
         \hline $e_{adc}$ \cite{jonsson2011empirical}& 0.25pJ \\
         \hline $e_{dac}$ \cite{palmers201010}& 0.01pJ\\
         \hline $e_{opt}$ [\cref{eq:e_opt}] & 0.01pJ \\
         \hline $e_{load}$ for 4$\mu m$ pitch, $N=256$  [\cref{eq:e_load}]& 0.08pJ \\
         \hline$e_{load}$ for 250$\mu m$ pitch, $N=40$ [\cref{eq:e_load}]& 0.8pJ \\
         \hline$e_{load}$ for 2.5$\mu m$ pitch, $N=2048$ [\cref{eq:e_load}]& 0.04pJ \\
         \hline 
    \end{tabular}
    \caption{Energy per operation for various operations of digital and analog computers.  These assume a technology node of 45nm and a voltage of 0.9V, and 8-bit values per operation.  The example of memory access energy assumes a bank size of 96kB, since this is the bank sized used to construct the TPU SRAM bank.}
    \label{tab:EnergyPerOperation}
\end{table}

\begin{figure*}
    \centering
    \includegraphics[width=0.7\textwidth,trim=0cm 4cm 6cm 
    0cm]{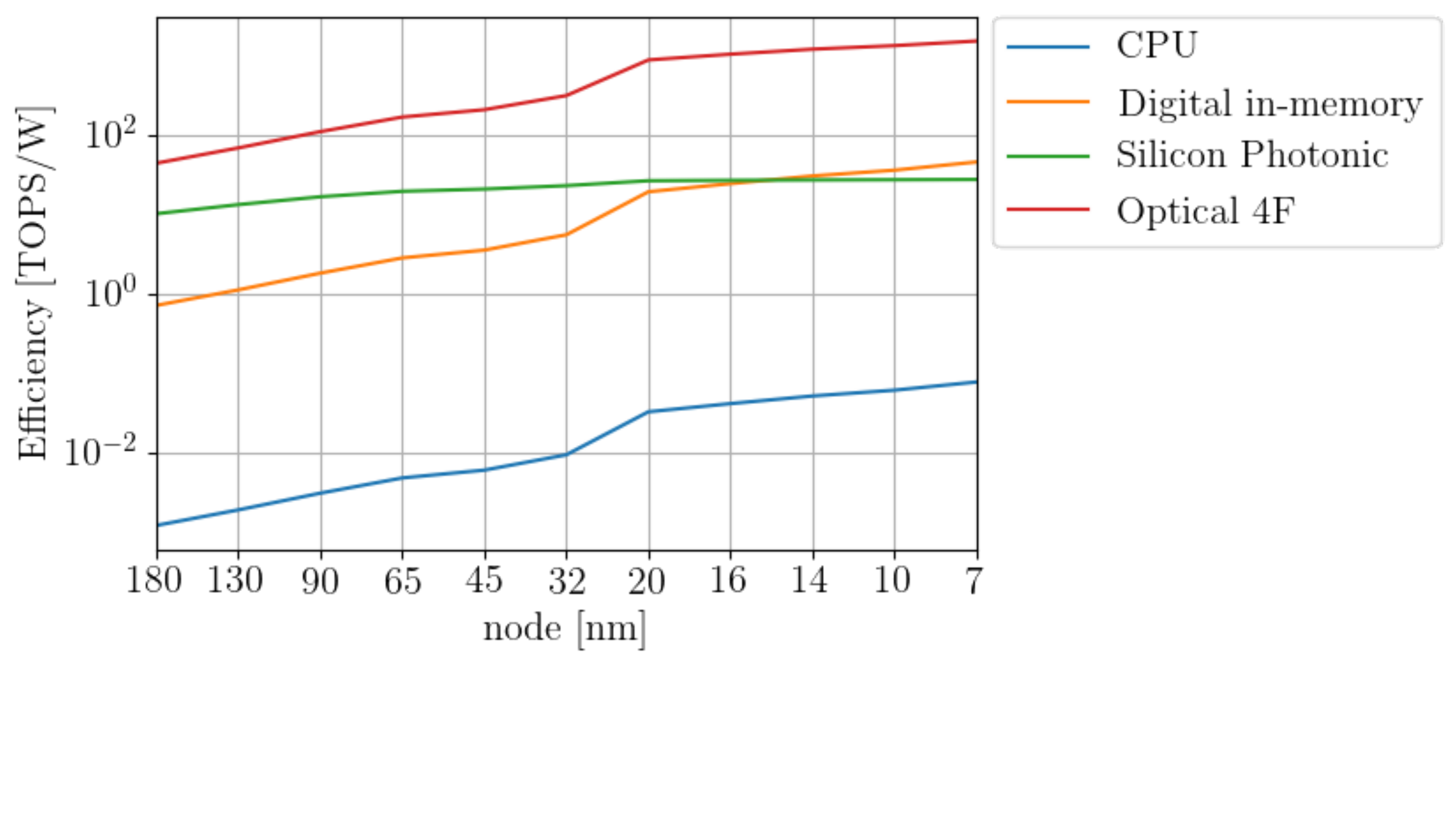}
    \caption{Efficiencies from analytic models of various compute architectures as a function of technology node.}
    \label{fig:AnalyticEfficiencies}
\end{figure*}

Each of these values depend on the CMOS technology node, but scaling laws can be used to interpolate between technology nodes \cite{stillmaker2017scaling}.  We compare the various compute platforms by considering a CNN layer with parameters given in \cref{tab:AnalyticParams}, and the resulting efficiencies are plotted as a function of technology node in \cref{fig:AnalyticEfficiencies}.

\begin{table}[]
    \centering
    \begin{tabular}{|c|c|c|}
        \hline Input Channels & $C_i$ &  128 \\
        \hline Output Channels & $C_{i+1}$ & 128 \\
        \hline Filter size & $k$ & 3 \\
        \hline Input size & $n$ & 512 \\
        \hline Arithmetic intensity & $a$ &  230 \\
        \hline
    \end{tabular}
    \caption{Convolution parameters used to estimate efficiencies of various processors in \cref{fig:AnalyticEfficiencies}.  The arithmetic intensity follows from the other parameters by \cref{eq:ConvArithmeticIntensity}.}
    \label{tab:AnalyticParams}
\end{table}

\begin{figure*}
    \centering
    \includegraphics[width=0.7\textwidth,trim=0cm 0cm 0cm 
    0cm]{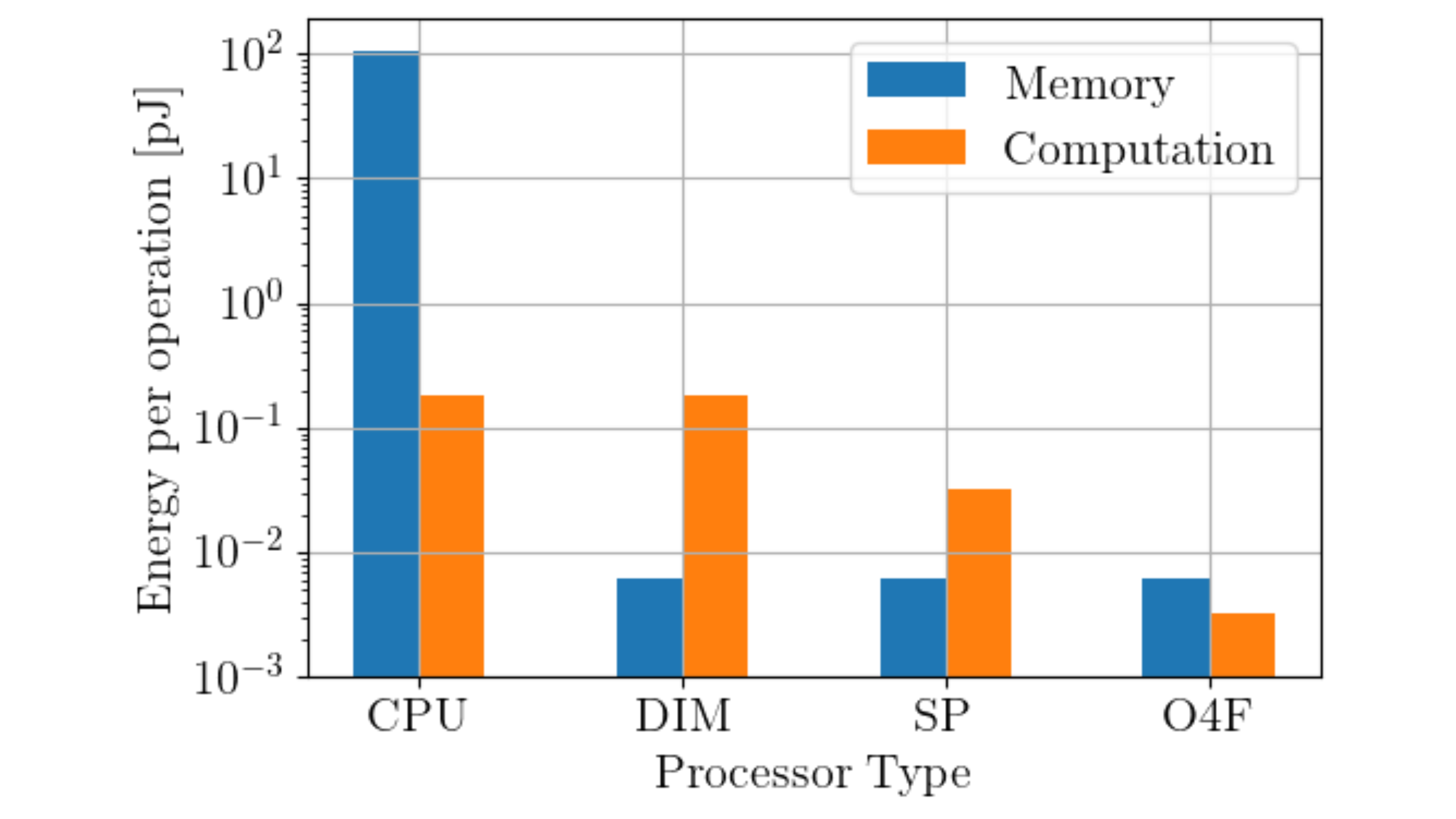}
    \caption{Contributions of energy consumption per operation for various processor types.  DIM is digital in-memory, SP is silicon photonic, and O4F is optical 4F system architectures.  The CNN layer parameters are in \cref{tab:AnalyticParams}, and assumptions about architectural details are given in the text.  The technology node is assumed to be 32nm for all processor types.}
    \label{fig:EfficiencyBreakdown}
\end{figure*}

While all processors improve with technology node, there is roughly an order of magnitude difference between digital in-memory compute processors and silicon photonic processors, and yet another order of magnitude difference to be expected between silicon photonic processors and optical 4F systems.  While this difference is clearly algorithm-dependent, the underlying hardware for analog compute systems must be large enough to be able to exploit the potential algorithmic advantages, which is what is enabled by moving from a two-dimensional silicon photonic processor to a fundamentally three-dimensional processor akin to an optical 4F system.

The breakdown of improvements into memory and computational energy reductions are shown in \cref{fig:EfficiencyBreakdown}, which shows the contribution to the energy per operation from memory and computational elements separately for each processor type.  Exploiting high arithmetic intensity with in-memory compute vastly improves energy consumption between CPUs and the other platforms by first reducing memory energy well below computational energy.  The analog processors in turn have reduced computational energy, with less computational energy on a per-operation basis for analog processors with more inputs.

It is worthwhile noting that the efficiencies reported in \cref{fig:AnalyticEfficiencies} for the digital in-memory processor are significantly higher than the Google TPU, which reported 0.3-2 TOPS/W depending on the CNN architecture, for a chip manufactured at a 28-nm node.  The in-memory compute digital processor modeled here has the same architectural parameters as the TPU: a 256 by 256 systolic array, and 24-MiB of SRAM divided into 256, 96-KB banks.  Here we predict that number should be roughly 5 TOPS/W, which is a significantly higher efficiency than reported in the literature \cite{jouppi2017datacenter}.  However, we note that this estimation simplifies the energy costs associated with the digital multiplication and storing and transporting data in and between each processing element in the systolic array.

The silicon photonics processor modelled in \cref{fig:AnalyticEfficiencies,fig:EfficiencyBreakdown} assumed an array size of 40 by 40, which is typical for most processors reported in the literature \cite{feldmann2021parallel,ong2020photonic,fang2019design,shen2017deep}, since the various modulator technologies typically require the array to have pitches in the 100-400 {\textmu}m range.  The computational energy consumption is highly limited by the optical modulator technology, which currently stands at around 7 pJ/byte, as discussed in section \ref{sec:SiliconPhotonicsDetails}.  We assume in our model that this will be improved to 0.5 pJ over time, but even with this assumed advantage it is clear in \cref{fig:AnalyticEfficiencies} that silicon photonics will have a difficult time maintaining an efficiency advantage over digital compute in memory technologies unless it is possible to scale up the processor sizes.  We also assume a 24-MiB SRAM for the silicon photonics processor, divided into 40, 600-KB SRAM banks, following the TPU architecture.

The optical 4F system is based on the architecture in \cref{fig:In-Reflection2}, with 4-Mpx SLMs and a 24-MiB SRAM divided into 2048, 12-KB banks, again following the TPU architecture.  The SLM pitch for DAC loads involved in active matrix addressing of the SLMs was assumed to be 2.5 {\textmu}m, which results in a line capacitance of 0.9 fF and a load energy of 40 fJ as shown in \cref{tab:EnergyPerOperation}.  The optical energy per pixel is based on 1550-nm light, and contributes 10 fJ/pixel per operation as shown in \cref{tab:AnalyticParams}.  The large array sizes enabled by realistic SLM dimensions are able to reduce computational energy consumption even below the memory consumption in \cref{fig:EfficiencyBreakdown}.

\section{Computational results}
Thus far, we have provided simple analytic formula that estimate the efficiency of various AI inference platforms on the basis of how they scale.  These formula are approximations with several limitations, the biggest of which is that they don't take into account situations where the matrices involved are too large for either the capacity of the in-memory compute device or its inputs.  In that circumstance the problem needs to be broken up into several smaller matrix multiplications.  In order to get around this limitation we developed cycle-accurate models of a systolic array and of an in-reflection optical 4F system, and tested those models when evaluating various CNNs for a given input image size.  The more accurate computational results are then compared with the analytic models from the previous sections.

\subsection{Systolic array efficiency estimation}

For analyzing the energy efficiency of a systolic array, we considered an architecture similar to that of the Google TPU [ref:TPU], with a weight-stationary systolic array of size 256 x 256 tiles.
Each of the 256 ports of the array has access to an individual 96-KB SRAM block, totaling 24 MiB of buffer memory for storing activations (i.e. inputs/outputs of a convolutional layer).
The weights are stored in DRAM and accessed based to the convolutional layer being executed.
The activations and weights are 8-bit fixed point.

In terms of energy costs, we used as reference the SRAM and MAC energy values for a 45-nm process at 0.9 V from \cite{horowitz20141}: SRAM read/write of 1.25 pJ/byte (8-KB memory) and 8-bit MAC operation of 0.23 pJ/byte.
To align with the SRAM block size of 96 KB in the TPU, the 8-KB SRAM energy cost was scaled in size by a factor of $\sqrt{96\textrm{K}/8\textrm{K}}=3.46$ in accordance with \cref{eq:MemoryScaling}, resulting in 4.33 pJ/byte.
Associated to each MAC operation, we also included the energy costs of the load and of the memory read/write inside each array tile (to store/propagate the 8-bit input and 32-bit accumulation = 40 bits).
A load energy cost of 2.82 fJ/bit was computed using eq. \ref{eq:e_load}, where the distance between array tiles was approximated based on the 256 x 256 array area occupancy (24\%) of the entire TPU chip ($331$ mm$^2$), resulting in a distance of 34.8 um between tiles.
The internal array memory energy cost was obtained by scaling the 8-KB SRAM block to 40 bits, resulting in $1.25$ pJ/byte $\times\sqrt{5/8\textrm{K}} = 31.25$ fJ/byte.

\begin{figure}
    \centering
    \includegraphics[width=0.5\textwidth]{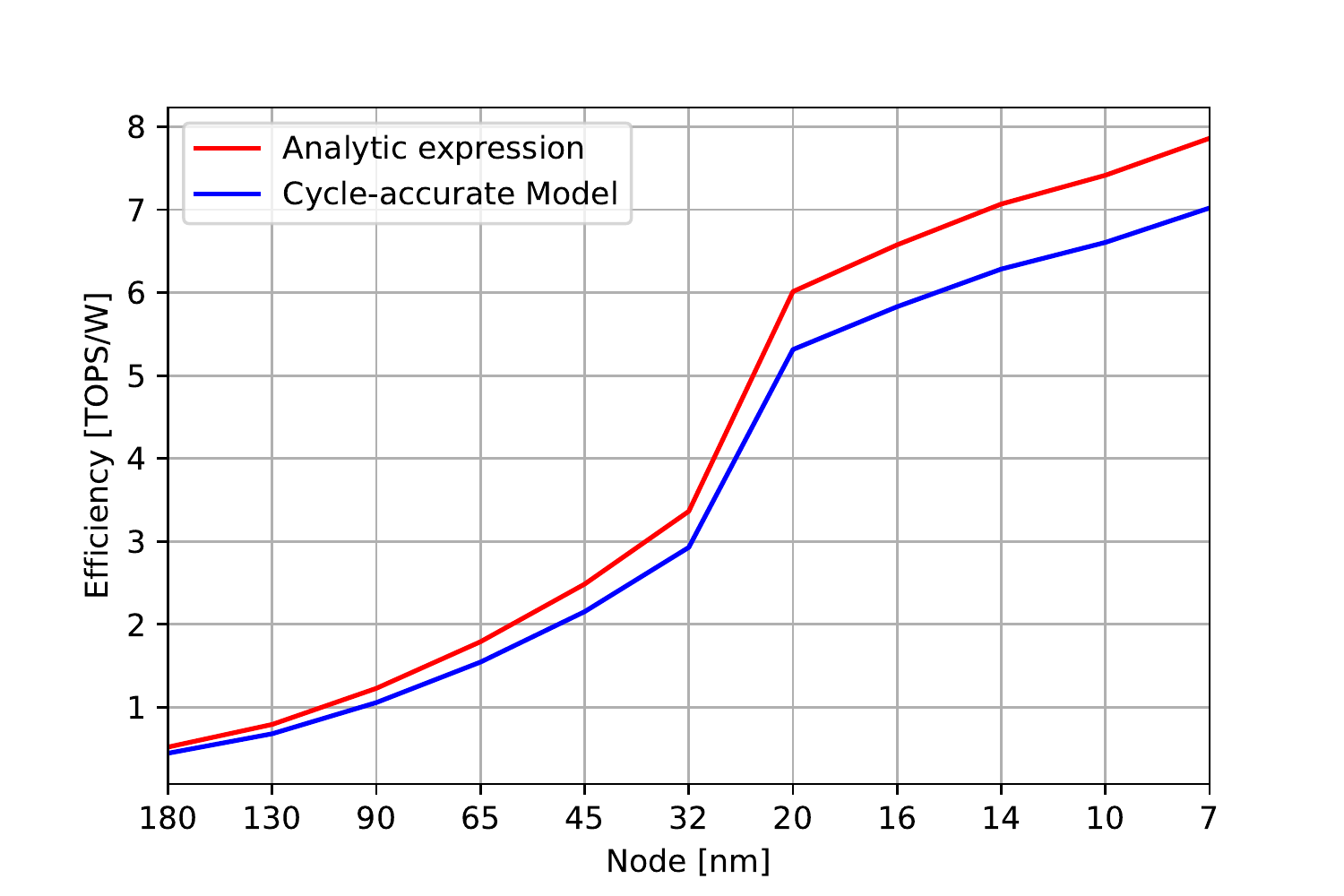}
    \caption{Efficiency comparison between a cycle-accurate model and the analytic expression given by \cref{eq:ConvArithmeticIntensity0} and the values in \cref{tab:CNN_Params}. Both models are running YOLOv3 (1-Mpixel input image) using a $256 \times 256$ weight-stationary systolic array and a 24-MiB SRAM (as in the Google TPUv1).}
    \label{fig:AnalyticVsComputational_TPU}
\end{figure}

Lastly, using the techniques presented in \cite{stillmaker2017scaling}, we scaled all the energy values (except for load, since it's not directly process-dependent) from 45-nm process to the appropriate technology nodes, ranging from 180 nm down to 7 nm. The results are presented in \cref{fig:AnalyticVsComputational_TPU}. Both the analytic expression and the cycle-accurate model follow the same trend, with a slight divergence as the technology node is reduced. This can be accounted for the fact that $e_{load}$ does not depend on technology node, and its cost starts becoming a dominating factor in the overall energy cost since the other energy sources diminish as node size reduces.

\subsection{Optical computer efficiency estimation}

For the optical 4F system we considered 4-Mpixel SLMs, along with the same 24-MiB SRAM as in the systolic array analysis. With this, the SRAM is partioned into 2048 equal parts (one per metasurface row), resulting in a size-scaled SRAM read/write energy of 1.55 pJ/byte. The DAC, ADC and laser energies were obtained using the values in \cref{tab:EnergyPerOperation} considering a 2.5-{\textmu}m pitch.

\begin{figure}
    \centering
    \includegraphics[width=0.5\textwidth]{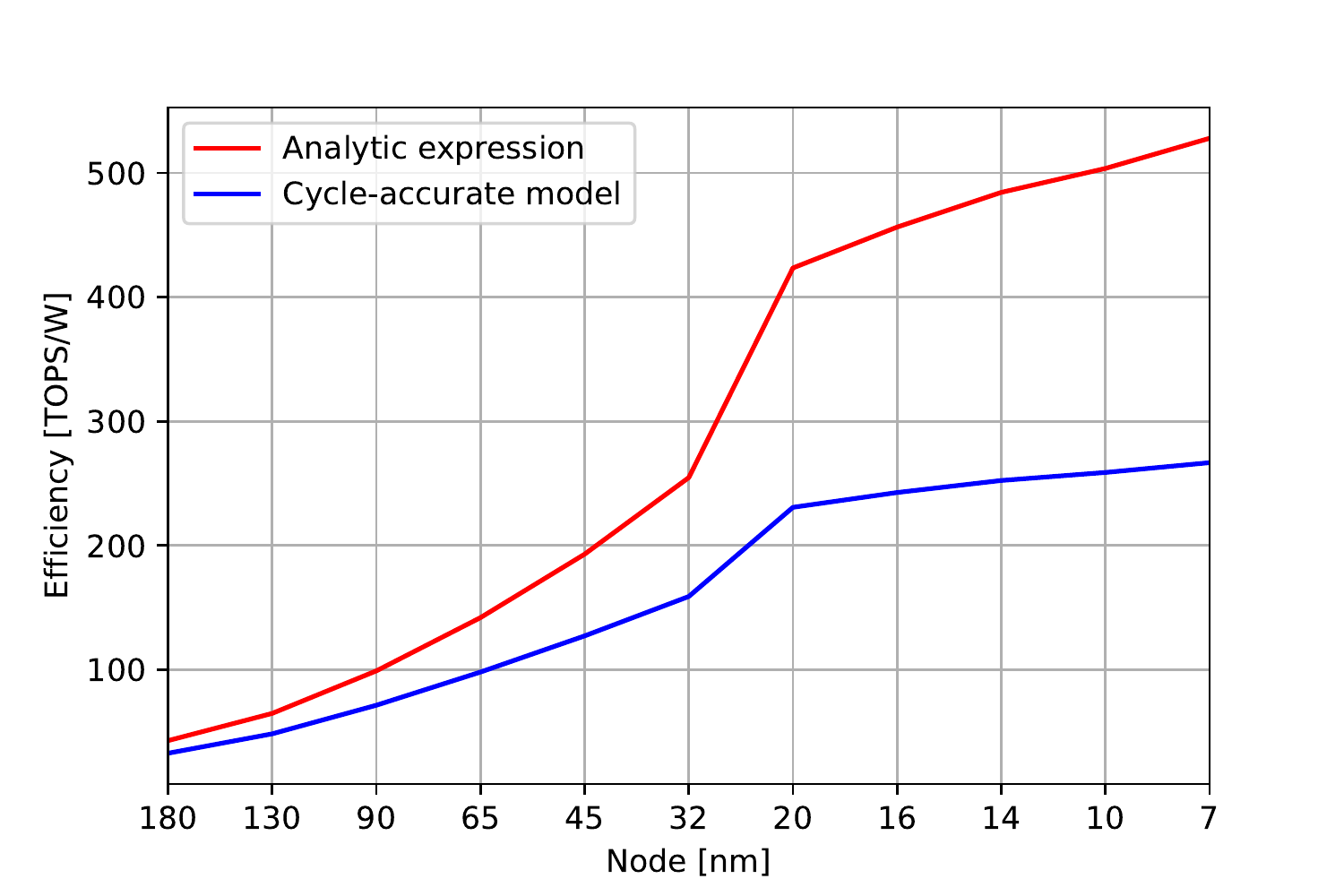}
    \caption{Comparison of \cref{eq:4F_Efficiency} with a cycle-accurate model of the optical 4F processor running YOLOv3 (1-Mpixel input image) using 4-Mpixel SLMs and a 24-MiB SRAM.}
    \label{fig:AnalyticVsComputational_OCU}
\end{figure}

A comparison between the analytic expression and a cycle-accurate model of the optical 4F system is presented in \cref{fig:AnalyticVsComputational_OCU}.
The figure provides an overall curve for the efficiency, with significant gain when constructing the device with smaller technology nodes.
The main differences which explain the divergence between the analytic and cycle-accurate models include:
\begin{itemize}
    \item The cycle-accurate model considers the exact number of metasurface executions to account for output detector ADC read operations, output memory accesses, and total laser energy consumed.
    \item \Cref{eq:4F_Dimensions} considers that the dimensions of the output are the same as that of the input (i.e. $m$=$n$), which naturally does not account for strides bigger than 1.
    \item The value of $e_{dac}$ in \cref{eq:4F_Efficiency} is comprised of $e_{dac,1} + e_{load} + e_{opt}$, resulting in an energy cost based on number of active pixels in the metasurface.
    However, the cycle-accurate model more precisely estimates the energy costs by separating the pixel-wise energy ($e_{dac,1} + e_{load}$) from the metasurface size-dependent laser energy ($e_{opt}$).
\end{itemize}

\subsection{Optical computer energy cost distribution}

\begin{figure*}
     \centering
     \begin{subfigure}[b]{0.45\textwidth}
         \centering
         \includegraphics[width=\textwidth]{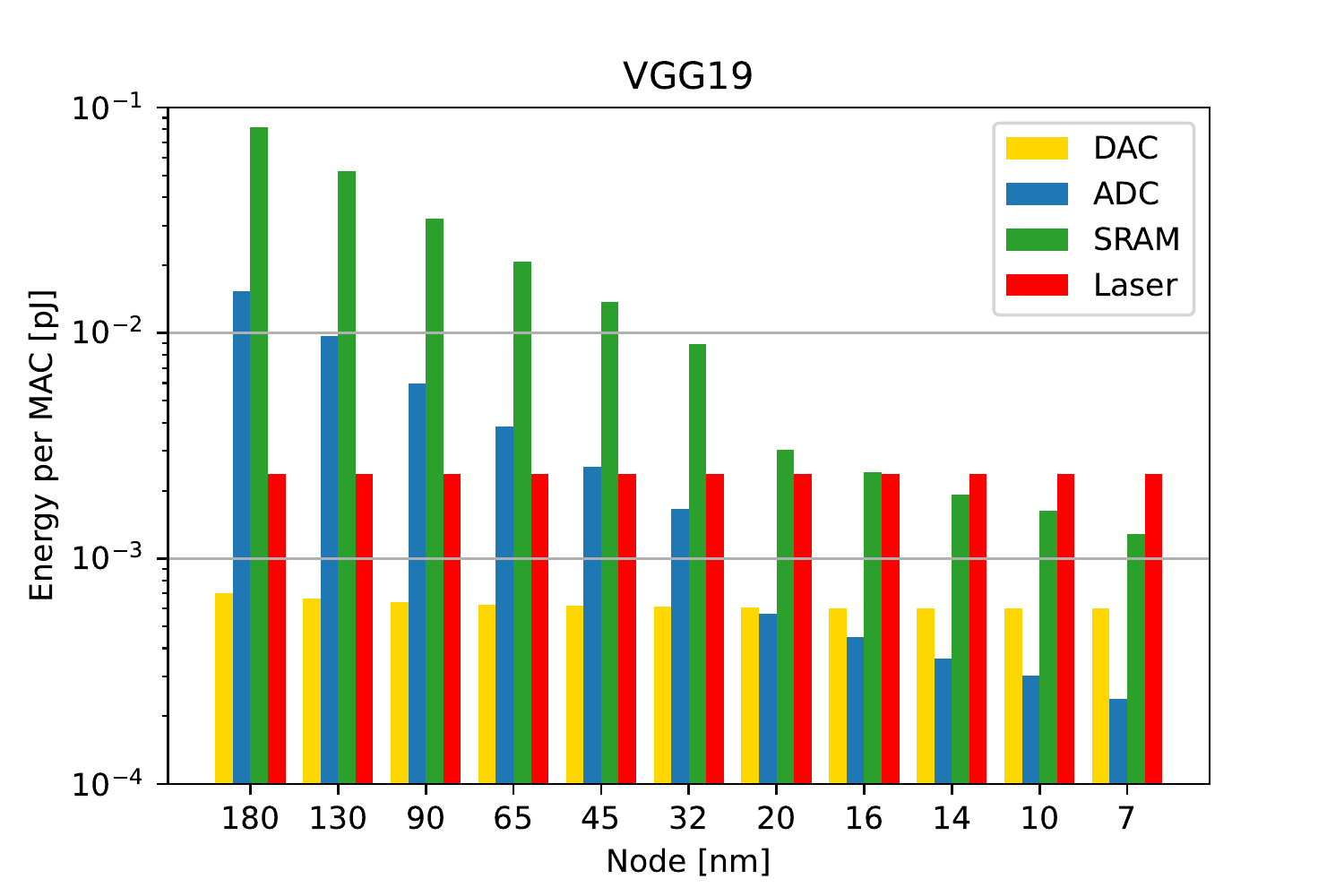}
     \end{subfigure}
     \hfill
     \begin{subfigure}[b]{0.45\textwidth}
         \centering
         \includegraphics[width=\textwidth]{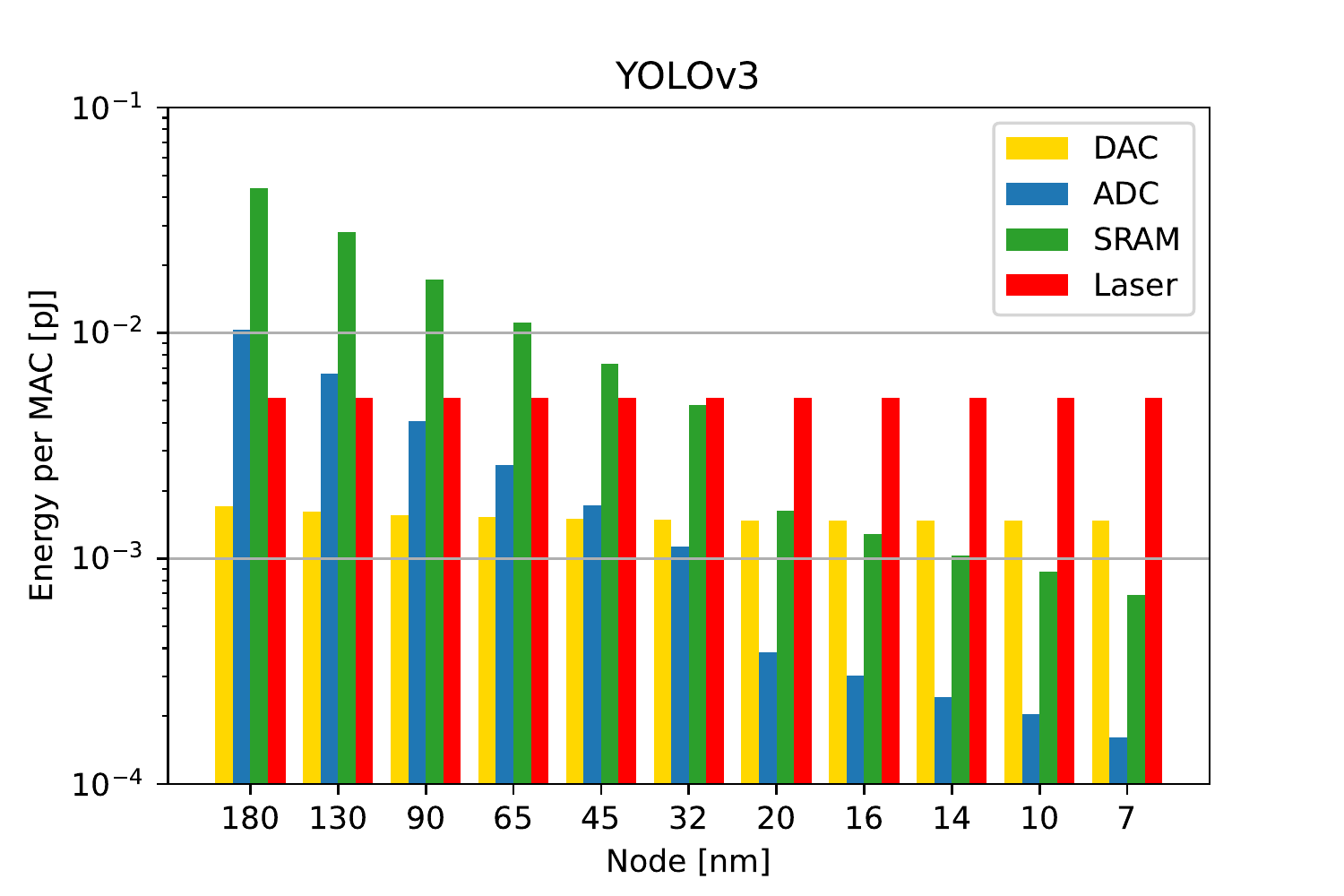}
     \end{subfigure}
    \caption{Energy cost distribution for the cycle-accurate model of the 4F optical system running VGG19 (left) and YOLOv3 (right).}
    \label{fig:OCUEnergyBars}
\end{figure*}

The cycle-accurate model for the optical 4F system can provide a detailed summary of the energy cost distribution based on four different system components: DAC, ADC, SRAM, and laser. 
These results for VGG19 and YOLOv3 across different technology nodes are presented in \cref{fig:OCUEnergyBars}, with the values specified in picojoules per MAC operation.

Naturally, as the node size reduces, ADC and SRAM energy costs decrease. On the other hand, the DAC energy includes the dominating $e_{load}$ in its composition, and since the latter is technology node-independent, we see very little reduction in the overall DAC energy cost throughout the different nodes. Just as with $e_{load}$, the laser energy $e_{opt}$ does not change with technology node and is, thus, constant.

Comparing the energy cost distributions between VGG19 (left) and YOLOv3 (right), it is curious to note that a network with a much larger arithmetic intensity as in the case of VGG19 (refer to \cref{tab:CNN_Params}) presents a higher SRAM energy per MAC operation. This can be explained by the fact that the cycle-accurate model takes into account the sizes of the SLMs and the inputs, making the VGG19 network slightly less efficient in terms of placement of the input image pixels onto the metasurface due to it presenting (on average per layer) larger input images with more channels. This results in more metasurface executions -- and, consequently, more output activation buffering (SRAM read/write) -- to complete the convolutions in the network. If we consider an infinitely large metasurface, then this artifact naturally goes away and VGG19 becomes more efficient than YOLOv3 in terms of SRAM energy per MAC operation.

\section{Conclusions}
In-memory compute and analog compute techniques are both effective techniques to address different contributions to total processor energy cost.  While in-memory compute is able to reduce memory access energy per operation in the context of a high arithmetic intensity algorithm, analog compute is able to reduce the computational energy itself in proportion to the scale of the analog processor. Convolutional neural networks are a perfect application for such analog, in-memory compute architectures since they have high arithmetic intensity, large linear operators, and typically require low bit precision for forward propagation.

To provide some intuition regarding how much energy efficiency can be improved using one or both of these techniques, we have provided simple analytic formula estimating the efficiency for a range of processor types, including digital in-memory processors like a systolic array, analog in-memory compute processors, and optical 4F systems which are a class of analog in-memory processor that are specialized to convolutions. These analytic formula, when applied to average neural network parameters provided in \cref{tab:CNN_Params,tab:LNM_Analog,tab:LNM_Analog} show good agreement with a cycle accurate model of the TPU architecture in \cref{fig:AnalyticVsComputational_TPU,fig:AnalyticVsComputational_OCU}.

As shown in \cref{fig:AnalyticEfficiencies} of these approaches perform orders of magnitude better than CPUs, at any modern technology node.  The largest improvement is due to the reduction in memory access energy per operation for in-memory compute processors, as shown in \cref{fig:EfficiencyBreakdown}.  However, this technique is so effective that computational energy becomes the dominant contribution, which is then improved by analog computing.  Since analog computing's energy advantage is proportional to the scale of the analog processor, optical 4F systems have a particular advantage in that they can be scaled large enough to reduce computational energy per operation below the minimum memory energy required for an in-memory compute processor when evaluating a modern CNN algorithm.

\appendix
\section{Processor Energy Model Parameters}
In this section, we provide derivations for the typical energies per operation associated with memory access, digital MAC, ADC, and DAC, which are all necessary in order to properly compare the various computing schemes discussed in this paper.

The energy consumed by a digital MAC operation will scale as the number of gates involved with the logical unit: indeed the lower bound of a digital MAC is set by the Landauer limit, which is proportional to the number of gates.  For a serial-parallel multiplier, the number of gates $G$ is $G=6B^2$, and for other  multiplier implementations, the area or gate-count is still proportional to $B^2$ \cite{ComputerArithmeticAlgorithms}, where $B$ is the number of bits of the operand.  A full adder has and additional nine gates per bit, so we can write
\begin{equation}
    e_{mac}=\gamma_{mac}(6B^2+9B)kT
\end{equation}
where $k$ is Boltzmann's constant, $T$ is temperature, and $\gamma_{mac}$ is a dimensionless constant.  Landauer's limit specifies that the energy per MAC is bounded on the lower end by $\gamma_{mac}>\mathrm{ln}(2)$.\cite{landauer1961irreversibility}  Typically, $\gamma_{mac}\approx122,500$ for a 45-nm process \cite{horowitz20141}, so current digital multipliers have several orders of magnitude improvement that could in theory be achieved.

All accelerators will need internal memory for both neural network parameters and intermediate variables (unless an analog processor is built with a large enough capacity to store the entire network which is currently impractical).  Digital electronic SRAM banks have an energy per operation that scales as the length of the bit and word lines used to address and write data to the SRAM, since most of the power is consumed in charging and discharging the effective capacitors formed by these lines.  In general then, the energy per memory access can be written as,\cite{horowitz20141}
\begin{equation}\label{eq:MemoryScaling}
    e_m=e_{m0}\sqrt{N_m}
\end{equation}
where $N_m$ is the size of the memory bank, and $e_{m0}$ is a constant with units of energy.  The scaling presented here is not reflective of the lower bound of energy consumption according to the Landauer limit, since currently it's the charging and discharging of capacitive lines that drives the energy consumption of SRAM rather than switching gates, which is why it scales according to the root of the array size.  In the limit of a single bit cell, one might compare $e_{m0}$ with the Landauer limit by setting $e_{m0}=\gamma_m kT$.  The resulting $\gamma_m$ is many more orders of magnitude away from the Landauer limit than even digital MACs: $\gamma_m\approx3\times10^6$ for a 45-nm CMOS process, which corresponds to an $e_{m0}\approx 5$ fJ.  It can be argued that both the sheer value of $e_{m0}$ compared to the Landauer limit and the fact that power consumption in the capacitance of the addressing lines in SRAM leads to the energy scaling proportional to the root of the array size are broadly the source of computing's most severe energy problem.\cite{horowitz20141}  Fortunately, in the case of specialized processors implementing operations with high arithmetic intensity like convolutions, we are able to significantly mitigate this problem.\cite{sebastian2020memory}

For analog computation, ADC energy depends exponentially on bit precision, most fundamentally because it requires sufficient signal to noise ratio to distinguish the levels.  When these levels are defined in terms of linear voltage steps, the ADC energy per sample is\cite{jonsson2011empirical,saberi2011analysis}
\begin{equation}
    e_{adc}=\gamma_{adc} kT2^{2B},
\end{equation}
where $k$ is Boltzmann's constant, $T$ is the temperature, and $\gamma$ is a dimensionless constant.  It has been argued\cite{jonsson2011empirical} that $\gamma_{adc}$ is bounded on the lower end at $\gamma_{adc}>3$ by thermal noise, and presents an empirical survey showing that the state of the art value for on-chip ADCs is $\gamma_{adc}\approx1404$ for a 65-nm process, which scales to about 927 at 45 nm.

DACs scale in the same manner as ADCs:
\begin{equation}\label{eq:DAC_Scaling}
    e_{dac}=\gamma_{dac} kT2^{2B}
\end{equation}
and for similar arguments.  However, the value for state-of-the art on $\gamma_{dac}$ is $\gamma_{dac}\approx39$.\cite{palmers201010}

However, the expression in \cref{eq:DAC_Scaling} only takes into account the power burned in the DAC circuitry itself, and not the power consumed driving the analog processor load.  For example, the load of the bitline associated with a ReRAM processor in \cref{fig:CIM_Analog}(b) will be very different from the load associated with the variable optical attenuator (VOA) in the optical analog processor in \cref{fig:CIM_Analog}(b).  An optical processor will have an additional energy contribution from the optical laser power itself, which can be considered effectively part of the load energy.  Therefore we can write,
\begin{equation}
    e_{dac,i}=\gamma_{dac}kT2^{2B}+e_{load,i}
\end{equation}
for both $e_{dac,1}$ and $e_{dac,2}$.   In the following subsections these quantities are estimated for both analog, memristive processors and silicon photonic processors.

However, we note that for physically large arrays the load can often be dominated by the capacitance of the row and column addressing lines.  The formula for the energy dissipation due to the capacitance of the bitlines and wordlines is,
\begin{equation}
e_{load,i}=(1/2)\mathcal{C}LV^2
\label{eq:e_load}\end{equation}
where $\mathcal{C}$ is the capacitance per unit length of the line, and $L$ is the line length.  For reference, a typical CMOS copper trace has a capacitance of around 0.2 fF$/\mu m$\cite{weste2015cmos}, so for a process with 0.9 volts they typically consume 0.08 fJ$/\mu m$ per operation.

\subsection{Silicon Photonics Analog Processors}
\label{sec:SiliconPhotonicsDetails}
For an optical computer, there is both an optical and an electrical component to the load for the driving of the inputs:
\begin{equation}
    e_{load,1}=e_{elec}+e_{opt}.
\end{equation}

The electrical component will involve driving some kind of electro-optic modulator, and the energy per operation will depend on the capacitance of that component in the usual way.  In the context of silicon photonics, this might be a variable optical attenuator (VOA) on the data input, while a mach-zender interferometer, MEMS modulator, or phase-change modulator is often used to store the weight data in the array.  Some of the lower energy approaches used for electro-optic modulators are plasmonic resonators.  Out of these, the lowest recorded to date energy per modulation of plasmonic modulators is around $e_{elec}\approx 9$ pJ.\cite{dabos2022neuromorphic,haffner2018low}  These are comparable to electro-optic modulators made of doped silicon micro-ring resonators tuned via carrier plasma dispersion have been demonstrated with roughly 0.9pJ/bit, or ~7pJ/B.\cite{sun2015single}  It may be possible to design optical modulators in the future with lower energy per sample than these figures.\cite{miller2017attojoule}

The optical contribution to the load itself will depend exponentially on bit precision since the dominant source of optical noise is shot noise.  Therefore for the optical component we can write, 
\begin{equation}\label{eq:e_opt}
    e_{opt}=\frac{\hbar\omega}{\eta_{opt}} 2^{2B}\equiv\gamma_{opt}kT2^{2B}
\end{equation}
where $\hbar$ is Plank's constant in units of angular frequency, $\omega$ is the frequency of the light, and $\eta_{opt}$ is the efficiency of the optical system and photodetector.  Conveniently, the optical power consumption also scales as $2^{2B}$ since the dominant source of noise is typically shot noise. To provide numbers for context, for 1550-nm light and an optical efficiency of $80\%$, we have $\gamma_{opt}\approx39$, which corresponds to $e_{opt}\approx10$ fJ.  In light of the energy per sample associated with current electro-optic technology, the optical contribution to the energy is negligible.

The load associated with the reconfiguration of the weights, $e_{dac,2}$, only has an electrical component, which will involve both the electro-optic modulators and the electrical bitlines used to address the array.  Ultra-low energy electro-optic modulators typically also have small dimensions in order to minimize the capacitance, on the order of a few microns, which leads to an additional energy consumption of a few femtojoules per element in the length of the array.  This is also negligible compared to the energy associated with the electro-optic modulator itself.  Therefore, both $e_{dac,1}$ and $e_{dac,2}$ are dominated by the electro-optic modulator energy.

\subsection{Memristive Analog Processors}
In the ReRAM processor, the load has two contributions: the capacitance associated with the conductive lines in the array, and the dissipation of charge in the memristors.  The pitch of ReRAM arrays tends to be limited by the size of the transistor placed at each node, which means the array bitlines and wordlines are relatively short and have low capacitance.  Nonetheless energy consumption in large arrays can still be dominated by the capacitance, which is given by \cref{eq:e_load}.

On the other hand, in a ReRAM array the energy per operation consumed by the memristors themselves can also be quite high since the energy is proportional both the size of the array, and proportional to their average conductance, and the conductance is limited by the quantum conductance of $G_0=2e^2/h$, where $e$ is the charge of an electron and $h$ is Plank's constant.  The conductance of these elements is therefore limited in the range of $G=G_0$ to $G=G_02^B$ for $B$-bit precision elements.

Memristors are highly nonlinear elements, so the input data is usually supplied with pulse width modulation instead of changing the voltage.  Therefore the energy consumed by the entire array can be written as a sum over all the memristors
\begin{equation}
    \langle E_{ReRAM}\rangle=\delta t\sum_{i=1}^M\sum_{j=1}^N\langle G_{ij}\rangle\langle V_j^2\rangle.
\end{equation}
where $\delta t$ is the samplint period.  Using the nominal value of the conductances and voltages for each memristor, we can simplify the equation to
\begin{equation}
    \langle E_{ReRAM}\rangle = \delta tMN \langle G\rangle V_{rms}^2.
\end{equation}
In one action of the array the number of MAC operations is $MN$, so the average energy per operation consumed by the array is actually a constant and \emph{is not reduced by scaling up the array} in the case of a ReRAM array:
\begin{equation}
    e_{ReRAM}\equiv \frac{\langle E_{ReRAM}\rangle}{MN}=\langle G\rangle V_{rms}^2\delta t
\end{equation}
As noted above, the conductance of memristors is only well behave above quantum conductance.  Assuming a uniform distribution, the average value will be half the dynamic range, and therefore $\langle G\rangle=2^{B-1}G_0$.

The energy is proportional to the square of the voltage, so we assume  this is limited to maintain $B$ effective bits of precision relative to the Johnson-Nyquist thermal noise $V_{noise}$ limit in the memristors.  For a clock period of $dt$, the thermal noise is,
\begin{equation}
    V_{noise}^2=\frac{4kT}{G_0\delta t}.
\end{equation}
since the maximal noise is given by the minimum conductance.  Setting $V_{rms}^2=(3/2)2^{2B}V_{noise}^2$ as the minimal required voltage to maintain $B$ bits of accuracy, the mimimum energy absorbed by the memristor array per operation is,
\begin{equation}
    e_{ReRAM}=3kT2^{3B}.
\end{equation}

While this is the ideal solution, in practice there is a minimum voltage that can be applied that is typically much higher than the thermal noise limit, and is on the order of $V_{rms}\approx70$ mV.   Using this estimate and a sampling period of $\delta t=1$ ns, the energy per operation due to the memristors is $e_{ReRAM}\approx 0.05$ pJ, which is about five times lower than the energy per operation in commercial memristor arrays, but nonetheless places an upper bound on the efficiency at $\eta\approx 20$ TOPS/W.

\begin{table}[]
    \centering
    \begin{tabular}{|c|c|}
         \hline Active ReRAM\cite{xia2019memristive,khakifirooz2012extremely} &  1-4$\mu m$\\
         \hline Optical phase-change material \cite{feldmann2021parallel} & 250 $\mu m$ \\
         \hline Optical Mach-Zender interferometer (MZI)\cite{shen2017deep} & 100 $\mu m$ \\
        \hline
    \end{tabular}
    \caption{Typical pitches for various analog compute modulation technologies.}
    \label{tab:my_label}
\end{table}

\begin{table}[]
    \centering
    \begin{tabular}{|c|c|}
         \hline $\gamma_m$ &  $3\times10^6$\\
         \hline $\gamma_{mac}$ & $1.2\times10^5$\\
         \hline $\gamma_{adc}$ & 583 \\
         \hline $\gamma_{dac}$ & 39 \\
         \hline $\gamma_{opt}$ & 105 \\
         \hline
    \end{tabular}
    \caption{Values of dimensionless constants for various operations.  These assume a technology node of 45nm and a voltage of 0.9V.  Optical efficiency is assumed to be 50\% for $\gamma_{opt}$.}
    \label{tab:gammas}
\end{table}

\bibliography{CIM}

\end{document}